 \documentclass[natbib]{svjour3}            % onecolumn (standard format)
 \smartqed  				    % flush right qed marks
 \usepackage{graphicx}
 \usepackage{aps-bibstyle}                  % use this style if not BibTeX.
 \journalname{SSRv}
 \def\ang{\AA}
 
 \def\gapprox{\lower.4ex\hbox{$\;\buildrel >\over{\scriptstyle\sim}\;$}}
 \def\lapprox{\lower.4ex\hbox{$\;\buildrel <\over{\scriptstyle\sim}\;$}}
 \newcommand{\apj}{{Astrophys. J.}}
 \newcommand{\SP}{{Solar Phys.}}
\begin{document}

\title{		GeV Particle Acceleration in Solar Flares and 
		Ground Level Enhancement (GLE) Events }

\titlerunning{Flare Acceleration and GLEs}        

\author{Markus J. Aschwanden$^1$}

\authorrunning{Aschwanden}

\institute{$^1$) Solar \& Astrophysics Laboratory, Lockheed Martin Advanced
		Technology Center, Bldg. 252, Org. ADBS, 3251 Hanover St.,
		Palo Alto, CA 94304, USA \\
              	Tel.: +650-424-4001\\
              	Fax: +650-424-3994\\
              	\email{aschwanden@lmsal.com}\\ }

\date{Received: 2010-Apr-30 / Accepted: ...}

\maketitle

\begin{abstract}
{\sl Ground Level Enhancement (GLE)} events represent the most energetic class 
of {\sl solar energetic particle (SEP)} events, requiring acceleration
processes to boost $\gapprox 1$ GeV ions in order to produce showers
of secondary particles in the Earth's atmosphere with sufficient intensity
to be detected by ground-level neutron monitors, above the background of
cosmics rays. Although the association of GLE events with both solar flares 
and coronal mass ejections (CMEs) is undisputed, the question arises about
the location of the responsible acceleration site: coronal flare reconnection
sites, coronal CME shocks, or interplanetary shocks? 
To investigate the first possibility we 
explore the timing of GLE events with respect to hard X-ray production in 
solar flares, considering the height and magnetic topology of flares, 
the role of extended acceleration, and particle trapping.  We find that 
50\% (6 out of 12) of recent (non-occulted) GLE events are accelerated 
during the impulsive flare phase, while the remaining half are accelerated 
significantly later. It appears that the prompt GLE component, which is 
observed in virtually all GLE events according to a recent study by
Vashenyuk et al., is consistent with a flare origin in the lower corona, 
while the delayed gradual GLE component can be 
produced by both, either by extended acceleration and/or trapping in flare 
sites, or by particles accelerated in coronal and interplanetary shocks.

\end{abstract}

\section{		Introduction				}

A key aspect that motivated this review is the question whether 
{\sl ground level enhancement (GLE)} events, which apparently require
acceleration processes that produce $\gapprox 1$ GeV particles, originate
from flare regions in the solar corona or from shocks driven by
coronal mass ejections propagating through the corona and 
interplanetary space.  GLE events represent 
the largest {\sl solar energetic particle (SEP)} events that accelerate
GeV ions with sufficient intensity so that secondary particles are
detected by ground-level neutron monitors above the galactic cosmic-ray
background (Lopate 2006; Reames 2009b). A catalog of 70 GLE events, 
occurring during
the last six solar cycles from 1942 to 2006, has been compiled 
(Cliver et al.~1982; Cliver 2006), which serves as the primary database
of many GLE studies. So, GLE events are very rare, occurring only about
a dozen times per solar cycle, which averages to about one event per year.
While GLE events with 1 GeV energies represent the largest energies produced
inside our solar system, they are at the bottom of the cosmic ray spectrum,
which covers an energy range of $\approx 10^9-10^{21}$ eV, exhibiting 
a ``spectral knee'' between particles accelerated inside our galaxy 
($\approx 10^9-10^{16}$ eV) and in extragalactic sources 
($\approx 10^{16}-10^{21}$ eV).

While coronal mass ejections (CMEs) are widely considered as the main 
drivers of geoeffective phenomena, as pointed out in the so-called
``solar flare myth'' paradigm (Gosling 1993), the acceleration site of 
high-energy particles detected in-situ in the heliosphere can often
not unambiguously be localized, and thus we have to consider both options. 
Several timing studies of GLE events have shown evidence for multiple
(impulsive and gradual) components, so there is often not a simple
dichotomy of acceleration sites.
In order to address our central question whether GeV particles producing
GLE events are either of solar or of heliospheric origin, we will review various
aspects of flare observations of GLE events, such as the relative timing 
of the release of GLE-associated particles during solar flares (Section 2.1), 
prompt flare-associated acceleration of GLE protons (Section 2.2),
the maximum energies of gamma-ray producing particles in solar spectra 
(Section 2.3), the height of acceleration regions (Section 2.4), 
the magnetic topology of solar flare acceleration regions (Section 2.5), 
and the role of extended acceleration phases and particle trapping in 
solar flares (Section 2.6). We summarize our conclusions in Section 3. 
Complementary aspects of the same question from the view of CME-associated 
shocks are treated in the article of Gang Li, and active region
characteristics of GLE-associated flares are reviewed by Nariaki Nitta
in this volume.

\begin{figure}
\centerline{\includegraphics[width=1.0\textwidth]{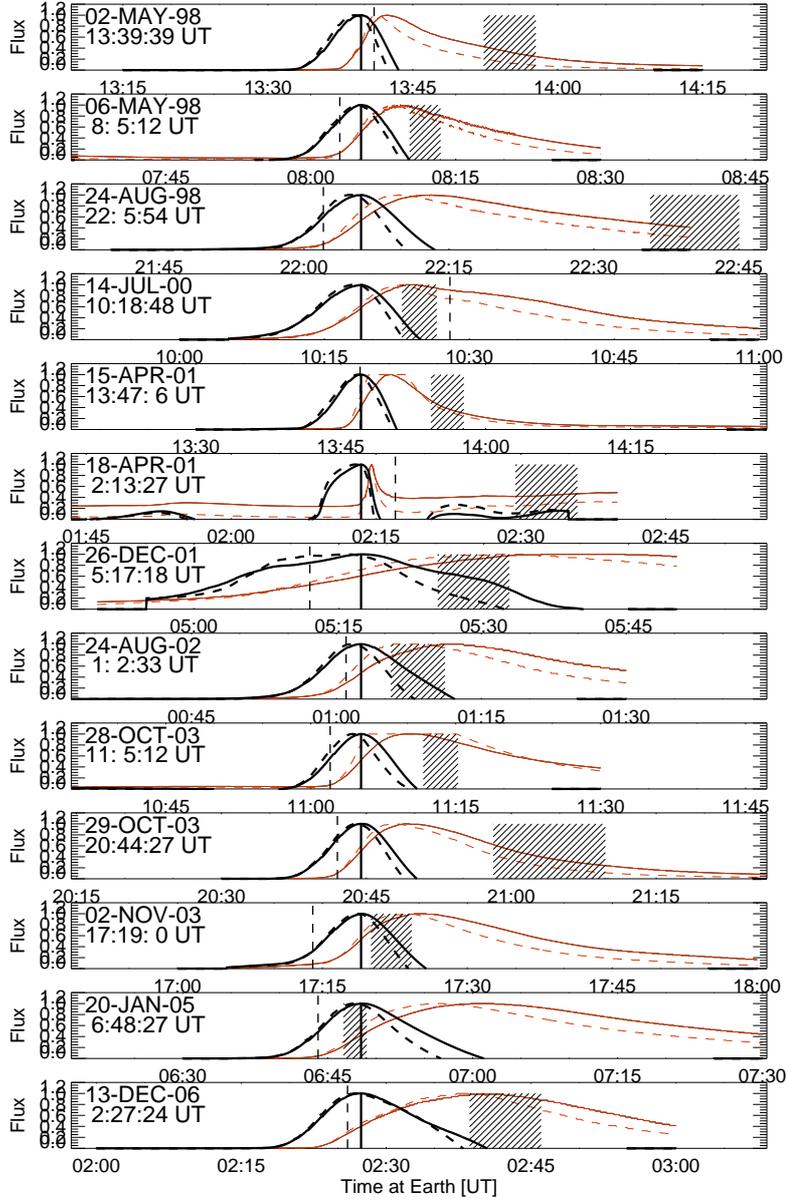}}
\caption{Time profiles of the GOES 1-8 \ang\ soft X-ray fluxes (solid red 
curve), GOES 0.5-4 \ang\ fluxes (dashed red curve), and their time derivatives
(thick black curves), which are a proxi for the hard X-ray flux, and solar 
particle release times with uncertainties (hatched areas) of 13 GLE events 
during 1998-2006. The time range is centered at the hard X-ray peak time 
(given on the left side). The {\sl solar particle release times (SPR)} are
determined from the velocity dispersion of particles detected with WIND 
by Reames (2009b), corrected by +8.3 min to the photon arrival time
at Earth. The start times of radio type II bursts are indicated with 
vertical dashed bars.}
\end{figure}

\section{	Flare Observations of GLE Events  	}

All GLE events are associated with solar flares of the most intense
category, i.e., GOES X-class flares in most cases, althouth there are
exceptions, e.g., see the 1979 August 21 event (Cliver et al.~1983) or
the 1981 May 10 event (Cliver 2006). At the same time,
{\sl coronal mass ejections} (CME) were reported in all recent cases. 
Thus we can say that flares and CMEs are both necessary conditions for 
a GLE event, but it leaves us with the ambiguity where the acceleration of
GeV particles reponsible for GLE events takes place. 
In the following we investigate and review various observational aspects
of relevant flare data that could shed some light into this question. 

\subsection{	Solar Particle Release Times	}

For the analysis of the timing of GLE events we are using initial solar 
release times that are normalized to the Earth-arrival times of electromagnetic
signals, based on the compilation of 16 events during 1997-2006 
(Gopalswamy et al.~2010; Reames 2009a) and of 30 events during 1973-2006 
(Reames 2009b).

\begin{table}
\caption{Timing of 13 GLE events during 1998-2006 (corrected to photon arrival time at Earth).}
\begin{tabular}{lllllrrl}
\hline
GLE	&Date	&Type II	&Hard X-ray	&SPR		&Delay	&Delay  &Location\\
\#	&	&start time	&peak time	&time		&SPR-II	&SPR-HXR&        \\
	&	&$t_{II}$ [UT]	&$t_{HXR}$ [UT]	&$t_{SPR}$ [UT] &[min]	&[min]	&        \\
\hline
56	&1998-May-02 & 13:41 & 13:39:39 & 13:55:02 & +14.0      & +15.4 & S15W15 \\
57	&1998-May-06 & 08:03 & 08:05:12 & 08:11:50 &  +8.8 	&  +6.6 & S15W64 \\
58	&1998-Aug-24 & 22:02 & 22:05:54 & 22:40:26 & +38.4 	& +34.5 & N35E09 \\
59	&2000-Jul-14 & 10:19 (10:28$^*$) & 10:18:45 & 10:24:50 &  -3.2 	&  +6.0 & N22W07 \\
60	&2001-Apr-15 & 13:47 & 13:47:06 & 13:54:02 &  +7.0 	&  +8.9 & S20W84 \\
61	&2001-Apr-18 & 02:17 & 02:13:27 & 02:32:38 & +15.6 	& +19.2 & $\approx$W115 \\
63	&2001-Dec-26 & 05:12 & 05:17:30 & 05:28:56 & +16.9 	& +11.4 & N08W54 \\
64	&2002-Aug-24 & 01:01 & 01:02:33 & 01:08:26 &  +7.4 	&  +5.8 & S02W81 \\
65	&2003-Oct-28 & 11:02 & 11:05:12 & 11:13:26 & +11.4 	&  +8.2 & S16E08 \\
66	&2003-Oct-29 & 20:42 & 20:44:27 & 21:03:56 & +21.9 	& +19.5 & S15W02 \\
67	&2003-Nov-02 & 17:14 & 17:19:00 & 17:22:08 &  +8.1 	&  +3.1 & S14W56 \\
69	&2005-Jan-20 & 06:44 & 06:48:27 & 06:47:50 &  +3.8 	&  -0.7 & N12W58 \\
70	&2006-Dec-13 & 02:26 & 02:27:24 & 02:42:20 & +16.3 	& +14.9 & S06W26 \\
\hline
\end{tabular}
{\it References: 
        Reames et al.~2001 (\#59);
	Aschwanden and Alexander 2001 (\#59); 
	Pohjolainen et al. 2001 (\#56);
	Shumilov et al.~2003 (\#56); 
	Vashenyk et al.~2003 (\#59); 
        Chertok et al.~2001 (\#59); 
	Torsti et al.~2004 (\#56); 
	Miroshnichenko 2005 (\#66,67); 
	Klassen et al.~2005 (\#65); 
	Simnett an Roelof 2005a,b (\#69); 
	Kocharov et al.~2006 (\#56); 
	Martirosyan and Chilingarian 2005 (\#69);
	Kuznetsov et al. 2006 (\#69); 
	Simnett 2006, 2007 (\#69); 
	Wang and Wang 2006 (\#69); 
	Li et al.~2007 (\#59); 
	Bieber et al. 2008 (\#69,70); 
	Bombardieri et al.~2008 (\#69); 
	Grechnev et al. 2008 (\#69); 
	McCracken et al. 2008 (\#69); 
	Moraal et al. 2008 (\#69);
	Watanabe et al. 2008 (\#65); 
 	Chupp and Ryan 2009 (\#65,69); 
	Li et al. 2009 (\#70); 
	Masson and Klein 2009 (\#69);
        Matthi\"a et al. 2009 (\#69);
        Reames et al. 2009a,b (all events); 
	Wang 2009 (\#59); 
        Vashenyuk et al. 2011 (\#59-61,65,67,69,70). --
	*) There are two type II events for the 2000-Jul-14 flare, 
	one starting at 10:19 UT (Chertok et al. 2001), the other
	starting at 10:28 UT (Gopalswamy 2010).}
\end{table}

In Fig.~1 we display the time profiles of 13 events (out of the 16 GLE
events occurring during 1994-2007) analyzed in Reames (2009a), for which
the solar release time could be determined with the velocity dispersion
method. The flare time profiles we show in Fig.~1 are the soft X-ray fluxes
$I_{SXR}(t)$ detected by GOES in the soft (1-8 \ang ) and hard (0.5-4 \ang ) 
channel (red solid and dashed curves), as well as their time derivatives
(black solid and dashed curves), which are generally considered as a good 
proxi of the hard X-ray flux $I_{HXR}(t)$,
$$
	I_{HXR}(t) \approx {dI_{SXR}(t) \over dt} \ ,
	\eqno(1)
$$
according to the Neupert effect (Neupert 1968; Hudson 1991; Dennis and
Zarro 1993; see also Section 13.5.5 in Aschwanden 2004 and references
therein). Note that the time derivatives of both soft X-ray channels
yield a near-simultaneous hard X-ray peak time ($<1$ min) 
within the accuracy needed here. The time profiles shown in Fig.~1 are
all of the same length (1 hr) and centered at the hard X-ray peak time
$t_{HXR}$, which we consider as the reference time for acceleration of the 
most energetic electrons during a flare. The physical explanation for the
Neupert effect is that the hard X-ray peak time coincides with the
time of most intense precipitation of accelerated high-energy electrons 
down to the chromosphere (within a time-of-flight interval of 
$\approx 10-100$ ms), which represents the time interval of most intense 
heating of chromospheric plasma, giving rise to the steepest increase of 
the soft X-ray flux during a flare. These hard X-ray peak times $t_{HXR}$ 
are listed in Table 1, which typically occur about $\approx 5-10$ min after
the soft X-ray onset times, but preceed the soft X-ray peak times
by about 1-15 min. The start or peak of soft X-ray emission (as listed
in GOES flare catalogs) should not be used as a reference time for
particle acceleration, because they rather bracket the beginning and
end of significant hard X-ray emission and the concomitant acceleration
phase.  

\begin{figure}
\centerline{\includegraphics[width=1.0\textwidth]{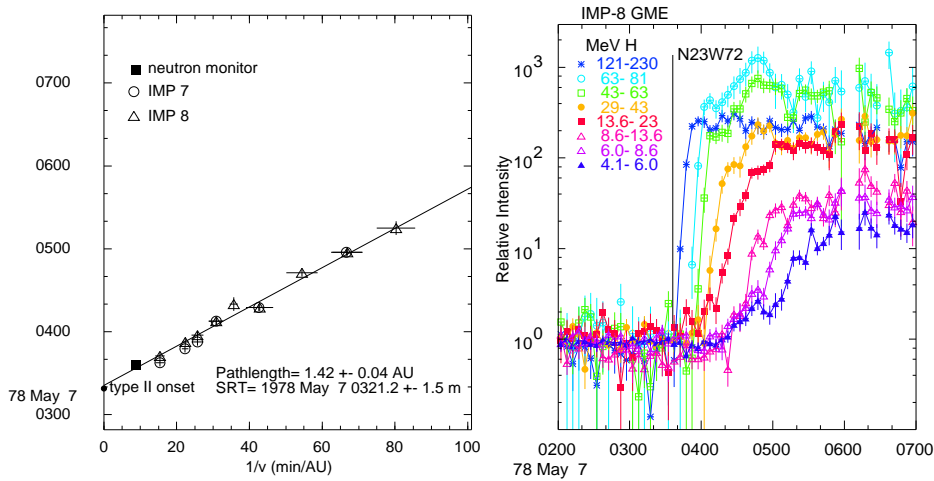}}
\caption{Example of a velocity dispersion measurement of GLE arrival
times of protons at IMP-7, IMP-8, or neutron monitors (y-axis in left panel) 
versus the reciprocal velocity $(1/v)$ (x-axis in left plot), which allows 
to extrapolate the energy release time $t_{SPR}$ at the source location 
($1/v \mapsto 0$), for the GLE event of 1978 May 7. The timing of the
relative intensity of the protons with different energies is shown
in the right-hand-side panel (Reames 2009b).}
\end{figure}

The {\sl solar particle release times} $t_{SPR}$ on the other hand,
were measured from the velocity dispersion of particles
arriving at ground-based neutron monitors or at particle detectors
in near-Earth satellites, such as with the {\sl Interplanetary Monitoring
Platform 8 (IMP-8)} and the {\sl WIND} spacecraft (Reames 2009a,b).
An example of such a measurement is shown in Fig.~2 for the GLE
event of 1978 May 7. 
A plot of the arrival time versus the reciprocal velocity $(1/v)$
shows a nearly-linear relationship that can be fitted with a
linear regression fit and yields the path length ($\approx 1.1-2.1$ AU)
as well as the so-called {\sl solar particle release time} $t_{SPR}$ 
within an accuracy of typically $\pm 1, ..., 3$ min. We list these
extrapolated solar particle release times $t_{SPR}$ determined by 
Reames (2009a) in Table 1 and mark their range of uncertainty with
a hatched area in Fig.~1. Defining the time interval of the impulsive
flare phase where the time derivative of the soft X-ray flux is positive
(corresponding to the time interval of impulsive hard X-ray emission
according to the Neupert effect), we find that 6 out of the 12 events
(50\%) have the earliest particle release times $t_{SPR}$ overlapping
with the impulsive hard X-ray flare phase (ignoring 
the occulted event \#61 on 2001-Apr-18), and thus are consistent
with flare-associated acceleration, which is generally called the 
{\sl prompt GLE component}. The remaining events have particle release times 
that start between 8 and 34 minutes later than the hard X-ray peak,
which could be produced by acceleration in CMEs, based on the relative timing
to hard X-rays. This results is not inconsistent with an earlier
review by Cliver et al.~(1982), where the most likely injection onset 
of GeV protons was found to conincide with the first significant
microwave peak (which is produced by gyrosynchrotron emission
and generally coincides with the production of nonthermal hard X-rays).

Another timing indicator is the onset time $t_{II}$ of metric 
type II bursts, which is also listed in Table I (from Gopalswamy et al.~2010).
The onset of type II bursts indicate the formation of a CME-related
shock front. These type II onset times $t_{II}$ are indicated with
a dashed vertical bar in Fig.~1, which seem to occur mostly during the
impulsive flare phase or rise time of soft X-ray emission.
Apparently, shocks in these intense X-class flares originate promptly 
during the flare, about at the same time when 
the magnetic field becomes stretched out below the erupting filament,
which triggers magnetic reconnection, particle acceleration, and 
chromospheric hard X-ray and soft X-ray emission subsequently. 
In half of the events, the solar particle release times $t_{SPR}$  
coincide with both the impulsive hard X-ray flare phase as well as
with the onset time $t_{II}$ of type II bursts (Fig.~1), and thus it is 
ambiguous whether the prompt component of GLE particles is accelerated in the
flare region or in a near-simultaneous type II or CME-associated shock front.

One GLE event, for which the timing was most extensively determined 
is the second-latest event of 2005 Jan 20. The steepest increase of the
soft X-ray flux occurs around 06:48 UT (Fig.~1). Hard X-ray time profiles 
from RHESSI show a peak of the 50-100 keV and 300-800 keV emission around 
06:45-06:46 UT, while the gamma-ray emission at 2-6 MeV, 23-40 MeV, and 
$>60$ MeV detected with SONG peak around 06:46-06:47 UT (Kuznetsov et al.~2006; 
Grechnev et al.~2008). The solar particle release time was determined
at 06:39 UT at the Sun (Reames 2009a), corresponding to a photon arrival time 
of 06:47 UT at Earth, which coincides with the hard X-ray 
and gamma-ray peaks, and about 3 min after the type II start time at 06:44 UT. 
Based on the temporal and spectral properties it was concluded that 
the acceleration site leading to the SEP/GLE spike is likely to be 
located in the flare region rather than in CME shocks, at least for
the leading SEP/GLE spike (Grechnev et al.~2008; Simnett and Roelof 2005a, 
2005b; Simnett 2006, 2007; Kuznetsov et al.~2006; Wang and Wang 2006;
Bombardieri et al.~2008; Chupp and Ryan 2009; Masson and Klein 2009), 
while a second component later on could be accelerated in a CME-associated 
shock at a distance of $\approx 3-5$ solar radii (McCracken et al.~2008; 
McCracken and Moraal 2008; Moraal et al.~2008). 

\subsection{	Prompt Flare-Associated Acceleration of GLE Protons  } 

Most GLE events exhibit a prompt component (PC) and a delayed component
(DC), which were identified in nearly all events in a recent study of
35 large GLE events during the period of 1956-2006 (Vashenyuk et al.~2011).
The prompt component prevails at the beginning of the event and is
characterized by an impulsive profile, strong anisotropy, and by an
exponential energy spectrum, i.e. $J(E) \propto \exp(-E/E_0)$ with
$E_0 \approx 0.5$ GeV (within a range of 0.3 GeV $\le E_0 \le$ 1.8 GeV). 
The delayed component dominates during
the maximum and decay phase of the events, has a gradual intensity
profile, a moderate anisotropy, and a powerlaw energy spectrum
(with a typical slope of $\delta \approx 5\pm1$).  
In comparison, the study of Moraal and McCracken (2011)
identifies a first (prompt) component in 14 events and a
secondary (delayed) component in 6 (out of 16 GLE) events 
during Cycle 23. There is discrepancy in 3 events, where 
Vashenyuk et al. (2011) registered a delayed component,
while Moraal and McCracken (2011) do not detect it.

Since CME-associated shocks last much later than the impulsive flare phase,
shock-accelerated particles are likely to increase in number and are
subject to a gradual release as long as the shock lasts, and thus
are likely not to have an impulsive time profile,
while flare-associated hard X-rays exhibit the same impulsive
time profile of particle acceleration naturally. 
However, some theoretical models of CME-driven shocks that include
the magnetic field gradient and the evolution of the shock-normal angle
from quasi-perpendicular to quasi-parallel orientation predict 
impulsive time profiles also (e.g., Zank, Rice, and Wu 2000).
Nevertheless, the fact that most GLE events (29 out of 35) analyzed 
in Vashenyuk et al.~(2011) exhibit a prompt component, togehter with 
our finding that the GLE start times occur during the impulsive hard 
X-ray phase in 50\%, is consistent with the interpretation of 
flare-associated acceleration for the prompt component.

\subsection{	Maximum Energies of Solar Gamma Rays 	}

There is no known high-energy cutoff of the electron bremsstrahlung spectrum;
the highest energies of observed bremsstrahlung are around several 100 MeV
(Forrest et al.~1985; Akimov et al.~1991, 1994a,b,c, 1996; 
Reames et al.~1992; Dingus et al.~1994; Trottet 1994; Kurt et al.~1996; 
Rank et al.~2001), see also reviews by Ramaty \& Mandzhavidze (1994)
or Chupp and Ryan (2009).
Gamma-rays were reported up to energies above 1 GeV with the {\sl Energetic 
Gamma-ray Experiment Telescope (EGRET)} on {\sl CGRO} during the 
GLE event \#51 on 1991 Jun 11 (Kanbach et al.~1992). The spectrum of
the flare (Fig.~3) could be fitted with a composite of a proton 
generated pion neutral spectrum and a primary electron bremsstrahlung 
component (Kanbach et al.~1992).
The GLE event \#69 on 2005 Jan 20 reported gamma rays, protons, 
and pion decay radiation up to energies $>200$ MeV (Grechnev et al.~2008).
In Table 2 we list the energy ranges of observed photons in hard X-rays
and gamma rays and the required primary particle energies for the various
high-energy processes operating in large solar flares (adapted from
Ramaty and Mandzhavidze 1994). While we have observational information of
maximum energies of electrons above 1 GeV, the maximum energies of protons
and ions in this regime can only be estimated indirectly, e.g., by the 
energy equi-partition argument, or by numerical simulations of a theoretical 
acceleration model.

The prompt component of GLE proton spectra were found to extend
to maximum energies in the range of 0.3 GeV $\le E_0 \le$ 1.8 GeV 
(Vashenyuk et al.~2011). Ground-based neutron detectors detect GLE protons 
at energies $E \gapprox 1.5$ GeV, which corresponds to a Lorentz factor of
$\gamma=E/(m_p c^2)=1.6$, a kinetic energy of $E_{kin}=m_p c^2 (\gamma -1)
=562$ MeV, and relativistic speed of $\beta=\sqrt{(1-1/\gamma^2)} 
\approx 0.78$.

\begin{table}
\caption{Hard X-ray and gamma-ray emission mechanisms operating in large
solar flares, with observed photon energy and primary particle energy
ranges (adapted from Ramaty and Mandzhavidze 1994)} 
\begin{tabular}{lll}
\hline
Process				&Observed photon	&Primary particle\\
				&energies		&energies	\\
\hline
Bremsstrahlung continuum	&20 keV$-1$1 MeV	&20 keV$-$1 MeV \\
				&$>$10 MeV		&10 MeV$-$1 GeV\\
Nuclear de-excitation lines	&0.4,...,6.1 MeV	&1-100 MeV/nucl.\\
Neutron capture line		&2.2 MeV		&1-100 MeV/nucl.\\
Positron annihilation radiation &0.511 MeV		&1-100 MeV/nucl.\\
Pion decay radiation		&10 MeV-3 GeV		&0.2-5 GeV	\\
Neutrons induced in atmospheric cascades&0.1-10 GeV	&0.1-10 GeV     \\
Neutron decay protons in space	&20-200 MeV		&20-400 MeV	\\
\hline
\end{tabular}
\end{table}

\begin{figure}
\centerline{\includegraphics[width=0.80\textwidth]{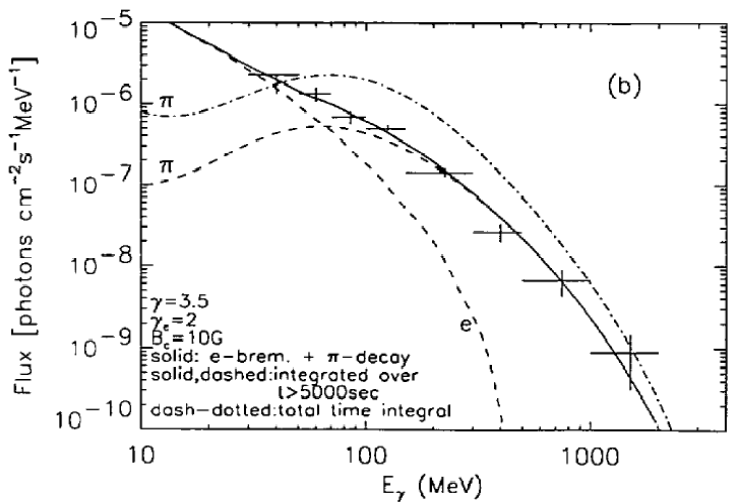}}
\caption{A gamma-ray spectrum observed with {\sl EGRET/CGRO} during
the GLE event \#51 on 1991 Jun 11, 02:04 UT flare, accumulated during 
03:26$-$06:00 UT (Kanbach et al.~1993). The spectrum is fitted with 
a combination of primary electron bremsstrahlung and pion-decay radiation. 
Note that pion decay is dominant at energies $\gapprox 40$ MeV
(Mandzhavidze \& Ramaty 1992).}
\end{figure}

Do these highest observed energies constrain or rule out any acceleration
mechanism? For DC electric field acceleration in sub-Dreicer fields,
the maximum velocity to which electrons can be accelerated is limited
by the value of the Dreicer field, which depends on the density and
temperature of the plasma, $E_D \approx 2 \times 10^{-10} n/T$
(statvolts cm$^{-1}$). Holman (1996) argues that
electron energies up to $10-100$ MeV can be attained for high densities
of $n_e \approx 10^{12}$ cm$^{-3}$ and low temperatures $T \approx 2$ MK
(yielding a Dreicer field of $E_D = 1 \times 10^{-4}$ statvolt cm$^{-1}$,
i.e., 3 V m$^{-1}$), if electrons are continuously accelerated over a 
current channel with a length of $L=10-100$ Mm. The requirement for such 
large-scale DC electric fields, however, conflicts with the observed 
time-of-flight delays of hard X-ray pulses (e.g., Aschwanden et al.~1996) 
and the short inductive switch on/off time scales required for the 
observed subsecond hard X-ray pulses. 
Alternatively, Litvinenko (1996) envisions super-Dreicer electric
fields, which can generate arbitrarily high maximum electron energies
within much smaller spatial scales, and thus can be consistent with the
time-of-flight delays of the observed hard X-ray pulses. The maximum energy
for relativistic electrons obtained over an acceleration time $t$ is 
approximately,
$$
        \varepsilon (t) = e E l(t) \approx e E c t \ .
	\eqno(5)
$$
Litvinenko (2003) identifies plausible physical conditions with 
super-Dreicer fields of order $E \gapprox 100$ V m$^{-1}$
in reconnecting current sheets that lead to electron acceleration with
gamma-ray energies of a few 10 MeV in electron-rich flares or to the
generation of protons with energies up to several GeV in large gradual
flares. A similar acceleration model was applied to the GLE event \#59
on 2000 Jul 14 (Li et al.~2007).
Also stochastic acceleration can generate 10 MeV electrons and
1 GeV protons (Miller et al.~1997), if a sufficiently high wave turbulence
level is assumed (which, however, cannot easily be constrained by
observations). Thus, the maximum observed gamma-ray energies imply 
only constraints for the acceleration mechanism of sub-Dreicer 
electric DC fields, but not for super-Dreicer DC electric field,
stochastic, or shock acceleration mechanisms.

\subsection{	Height of Acceleration Region 		}

Since we have a temporal coincidence of GLE particle acceleration
with respect to flare hard X-ray emission in at least 50\%, we turn
now to the question of the spatial localization of acceleration sources.

One method to estimate the height of the acceleration region of GLE 
particles (Reames 2009a,b) is based on an assumed height $h_{II}$ of 
the start of radio type II bursts, corrected for the propagation
delay of the CME shock front with speed $v_{CME}$ during the time
interval $\Delta t=t_{SPR}-t_{II}$ between the start $t_{II}$ of the 
radio type II burst and the solar particle release time $t_{SPR}$,
$$
	h_{SPR} = h_{II} + v_{CME} (t_{SPR}-t_{II}) \ ,
	\eqno(2)
$$
which is always higher than $h_{II}$, because $v_{CME} > 0$ and
$t_{SPR} > t_{II}$. Reames (2009a,b) assumed a nominal average
height of $h_{SPR}=1.5\pm0.5$ solar radii based on standard
coronal density models $n_e(h)$ and statistical start frequencies
of type II bursts around $\nu\approx 100$ MHz (Kundu 1965),
which corresponds to an electron density of $n_e\approx 10^8$ cm$^{-3}$ 
in the case of fundamental plasma emission, i.e., 
$f_{pe} \approx 8980 \sqrt{n_e}$. Taking the propagation delay
of $(t_{SPR}-t_{II}) \approx 5-15$ min and the CME speeds of
$v_{CME} \approx 1000-3000$ km s$^{-1}$ into account, Reames (2009a,b)
arrived at estimated heights of $h_{SPR}\approx 2-5$ solar radii,
with a possible dependence on the heliographic longitude. 
This method implicitly assumes that particles are accelerated in
CME shocks, whose location is entirely tied to the height and
propagation speed of the type II and CME shock front.

\begin{figure}
\centerline{\includegraphics[width=0.8\textwidth]{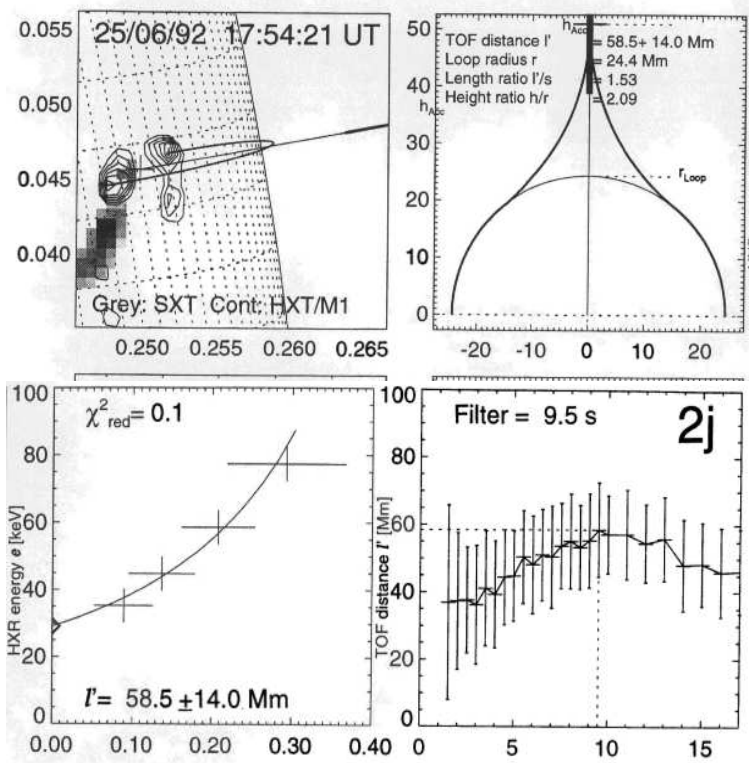}}
\caption{Altitude measurement of the acceleration source during the
1992 June 25 GLE precursor flare based on the velocity dispersion of 
electron time-of-flight delays, measured with the {\sl Compton Gamma Ray
Observatory (CGRO)}, {\sl Yohkoh/HXT}, and {\sl SXT}.
A hard X-ray image from HXT/M1 (23-33 keV) of the flare footpoint sources 
(contours) and a SXT soft X-ray image from Yohkoh/SXT (greyscale) are
shown in the top left panel. The energy-dependent hard X-ray time
delays in the range of $\Delta t \lapprox 300$ ms are shown in the
bottom left panel, yielding a time-of-flight distance of $L'=58.5\pm14.0$
Mm, which is projected onto the inferred loop geometry (top right
panel). The uncertainty of the TOF distance is derived from 
varying the filter time scale (bottom right panel), (Aschwanden et al.~1996).}
\end{figure} 

Alternatively, since half of the GLE events exhibit a starting time
$t_{SPR}$ during the impulsive phase of hard X-ray emission, we can estimate 
the height of their acceleration region from the hard and soft X-ray data.
The mildly relativistic electrons accelerated in a flare exhibit
a time-of-flight delay between their coronal acceleration site and
the chromospheric target region where bright hard X-ray emission is
observed, depending on their kinetic energy. From the velocity dispersion
of these energy-dependent hard X-ray time delays, the propagation distance 
and height of the acceleration region can be calculated, with proper 
correction for the
geometry of the trajectory and for the pitch angles of the particles. 

For a GLE precursor flare, which occurred on 1992 June 25, 17:32 UT 
(hard X-ray start time) at heliographic position N10/W75, such a 
time-of-flight measurement is available (Fig.~4), based on Yohkoh/HXT and SXT 
observations (Aschwanden et al.~1996). This flare is of the GOES
M1.4-class and peaked in soft X-rays at 17:54 UT. Concomitant
hard X-ray and radio emission is also studied in Wang et al.~(1995),
finding a spatial separation of 35 Mm between the simultaneous hard
X-ray and microwave emission, which provides also an approximate 
scale for the horizontal and vertical extent of the flare region. 
This flare is a precursor to a GLE
event 2 hrs later at the same location, which is listed as GLE event
\# 53 in Cliver (2006), peaking at 20:14 UT at position N10/W68, and is
classified as X3 GOES-class flare. We can consider the two rapidly 
following events as {\sl homologous flares} within the same active 
region that are likely to have a similar magnetic configuration. 
The time-of-flight distance for the precursor flare was evaluated 
from the time delay $\tau_{ij}$ between electrons with relativistic 
speeds of $\beta_i$ and $\beta_j$ in the hard X-ray photon energy 
range of $\epsilon \approx 20-80$ keV,
$$
	l_{TOF} = c \tau_{ij} \left({1 \over \beta_i}
	-{1\over \beta_j}\right)^{-1} \ ,
	\eqno(3)
$$
which after correction for electron pitch angle, helical twist of
field lines, and projection effects of loop size yielded a value of
$L'=58.5\pm14.0$ Mm, corresponding to a height of $h \approx 50 \pm 13$
Mm. This is a fairly typical height of the acceleration region for
large flares, amounting to about the double height of the soft X-ray
loops. From statistics of 42 flares, an average height ratio of
$h/h_{loop} \approx L/L_{loop} = 1.4\pm0.3$ was obtained (Aschwanden
et al.~1996), for flare loop radii of $r_{loop} \approx 2-20$ Mm.
Thus, the height range of acceleration regions in flares amounts to
$h \approx 4-40$ Mm, which corresponds to $\lapprox 5\%$ of a solar
radius. In summary, since about 50\% of the GLE events are consistent
with a particle release time during the flare hard X-ray phase,
they are expected to have acceleration heights of $h \lapprox 0.05$
solar radii. 

\begin{figure}
\centerline{\includegraphics[width=1.0\textwidth]{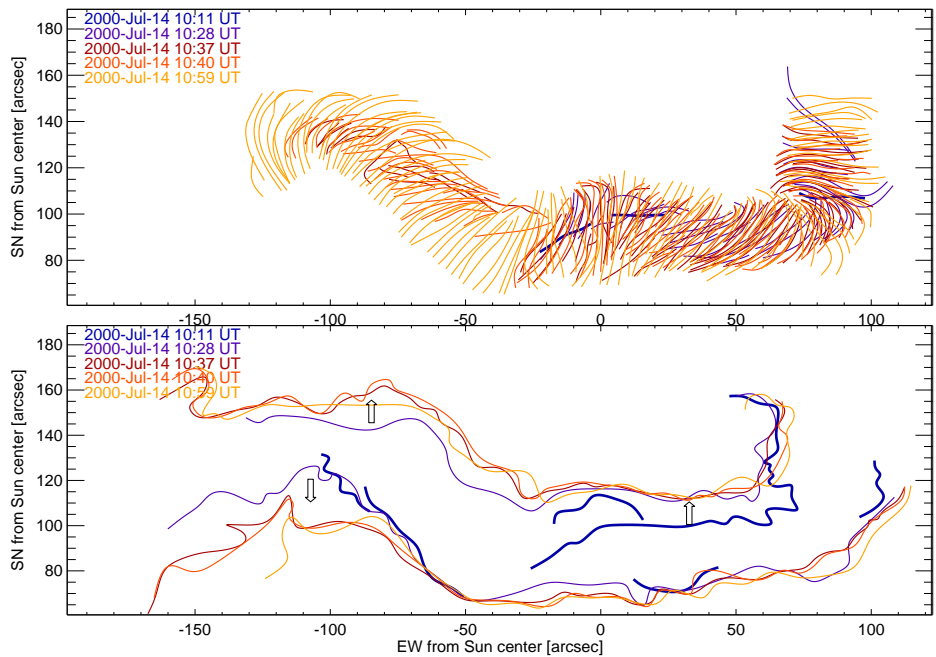}}
\caption{{\sl Top:} tracings of individual flare loops from {\sl TRACE} 171
\ang\ images of the Bastille-Day flare 2000-Jul-14. The five sets of loops 
traced at five different times are marked with different greytones. 
Note the evolution from highly sheared to less sheared loops. 
{\sl Bottom:} the position of the two flare ribbons traced from 171 \ang\
images. Note the increasing footpoint separation with time 
(Aschwanden 2002).}
\end{figure}

\begin{figure}
\centerline{\includegraphics[width=1.0\textwidth]{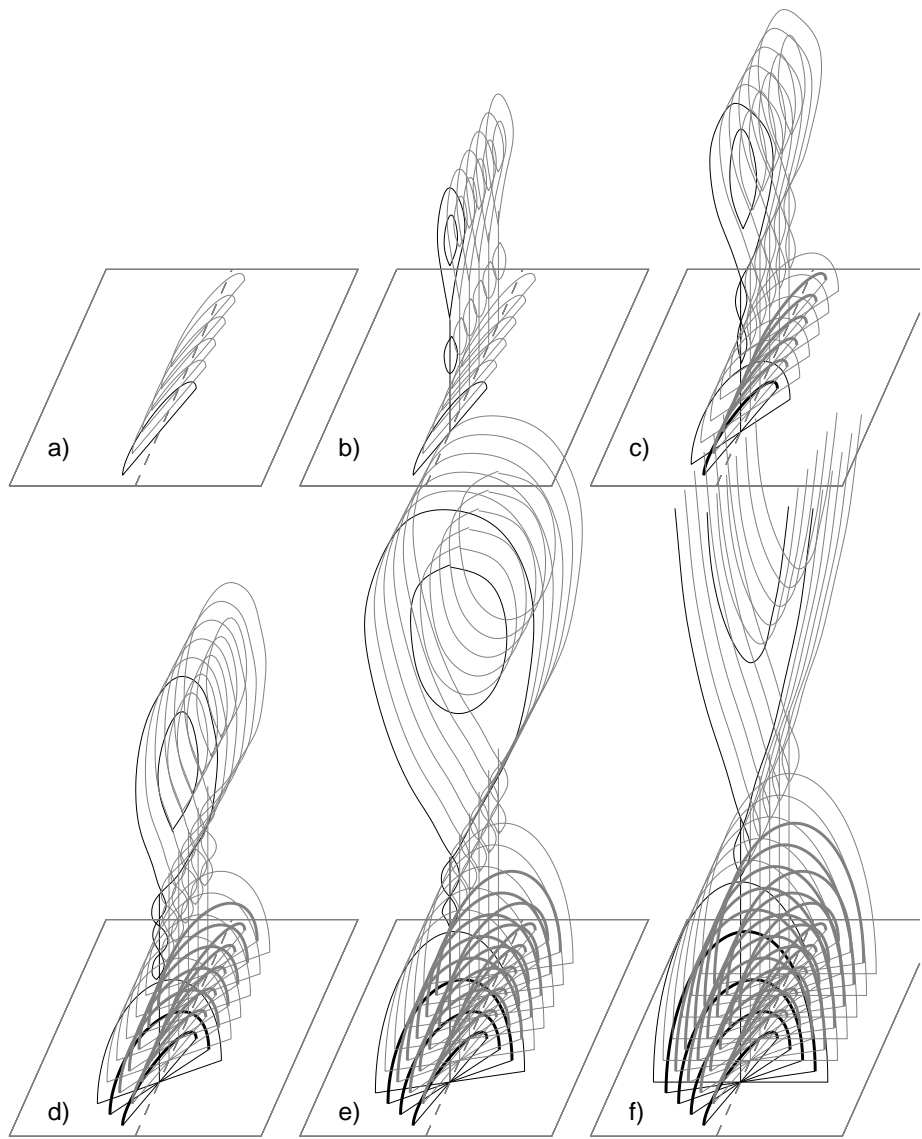}}
\caption{Scenario of the dynamic evolution during the Bastille-Day 
2000-Jul-14 flare: {\sl (a)} low-lying, highly sheared loops above 
the neutral line become unstable first; {\sl (b)} after loss of 
magnetic equilibrium the filament jumps upward and forms a
current sheet according to the model by Forbes and Priest (1995). 
When the current sheet becomes stretched, magnetic islands form 
and coalescence of islands occurs at locations of enhanced resistivity, 
initiating particle acceleration and plasma heating; {\sl (c)} the 
lowest lying loops relax after reconnection and become filled due to
chromospheric evaporation (loops with thick linestyle);
{\sl (d)} reconnection proceeds upward and involves higher lying, 
less-sheared loops; {\sl (e)} the arcade gradually fills up with 
hot flare loops; {\sl (f)} the last reconnecting loops have no shear 
and are oriented perpendicular to the neutral line. At some point 
the filament disconnects completely from the flare arcade and escapes 
into interplanetary space (Aschwanden 2002).}
\end{figure}

\subsection{	Magnetic Topology of Acceleration Regions 	}

The geometry and magnetic topology of a solar flare region has been
studied in most detail for the GLE event of 2000 July 14, 10:18 UT,
an X5.7 GOES-class flare (e.g., Aulanier et al.~2000; 
Aschwanden and Alexander 2001; 
Aschwanden 2002; Yan and Huang 2003; Aschwanden and Aschwanden 2008).
The solar energy release times of the particles responsible for the
GLE spike during the Bastille-Day flare is well-synchronized with the 
rise phase of hard X-ray emission (Fig.~1, Table 1; Reames 2009b;
Li et al.~2007; Wang 2009).
The Bastille-Day flare was observed near disk center, which provided
an excellent view on the projected flare area, which consists of a
classical double-ribbon structure as it is typical for most large
flares. The double-ribbon structure straddles along a neutral line
from east to west, and spreads apart as a function of time
(Fig.~5, bottom), as expected in the {\sl Carmichael- Sturrock- Hirayama- 
Kopp- Pneuman (CSHKP)} standard flare model. While the CSHKP model
essentially describes the vertical evolution in a 2-D cross-section
of the flare arcade, the Bastille-Day flare in addition shows also
the horizontal projection and the 3-D evolution in detail. From
loop tracings at different times during the flare (Fig.~5, top)
it is evident that low-lying, highly-sheared loops over the neutral
line brighten first, triggering a sequence of flare loops that
progresses to higher-lying and less-sheared field lines above the
neutral line, until we see a final double-ribbon flare arcade at 
orthogonal angles to the neutral line, spanning over a width of 
$w \approx 50$ Mm in NS direction and a length of $l \approx 250$ Mm
in EW direction. From this time evolution we can reconstruct the
3-D geometry of the magnetic field lines that are involved in the 
flare as shown in Fig.~6. Since the bright EUV postflare loops 
outline the relaxed field line configuration after magnetic reconnection,
we can directly infer the size and height of the magnetic reconnection 
region, which, statistically, is located about a factor of 1.5 above
the soft X-ray and EUV postflare loops (Aschwanden et al.~1996), 
estimated to be in an altitude
of $h \approx w \approx 50$ Mm with a length of $l \approx 250$ Mm.
This height range of the magnetic reconnection region confines the overall
volume of the particle acceleration region, although a highly 
fragmentated structure with many magnetic islands is expected, caused
by the tearing-mode instability and bursty reconnection (e.g., Sturrock 1966; 
Karpen et al.~1995, 1998; Kliem et al.~2000; Shibata and Tanuma 2001).  

Some consequences of this magnetic topology for GLE events are:
(1) The vertical X-type reconnection geometry allows particle acceleration
in upward and downward direction in a quasi-symmetric fashion, so that
particles of almost equal energies can be accelerated in both directions;
(2) The magnetic field lines above the vertical current sheet of the
main reconnection regions are likely to be open, which allows escape
of accelerated particles into interplanetary space and along
Earth-connected magnetic field lines;
(3) The relatively low height in the solar corona ($h \lapprox 50$ Mm)
provides a large reservoir of thermal particles that can be accelerated 
in sufficient numbers that are necessary for GLE detection.   
The escape conditions of accelerated particles into interplanetary
space requires open magnetic field lines, which exist not only in 
coronal holes but also to a substantial fraction in active regions.
Schrijver and DeRosa (2003) found from potential-field extrapolations 
of the global magnetic field over the entire solar surface that the 
interplanetary magnetic field (IMF) originates typically in a dozen 
disjoint regions, around the solar cycle maximum. While active regions 
are often ignored as a source for the interplanetary magnetic field, 
Schrijver and DeRosa (2003) found that the fraction of the IMF that 
connects directly to magnetic plages of active regions increases from 
$\lapprox 10\%$ at cycle minimum up to $30-50\%$ at cycle maximum, 
with even direct connections between sunspots and the heliosphere. 
Evidence for open-field escape routes was demonstrated with magnetic
field modeling, for instance for GLE event \#70, 2006 Dec 13, 02:27 UT
(Li et al.~2009). 
In flaring regions that expel a CME, particles accelerated in the
reconnection region behind the rising filament may be trapped inside
the CME bubble and cannot directly escape into interplanetary space
(Reames 2002), but may contribute to a seed population for secondary
acceleration in CME-associated shocks (McCracken and Moraal 2008). 
The two populations of directly-escaping and trapped-plus-accelerated 
particles may be distinguishable as impulsive and gradual phases of 
SEP events.

\begin{figure}
\centerline{\includegraphics[width=0.70\textwidth]{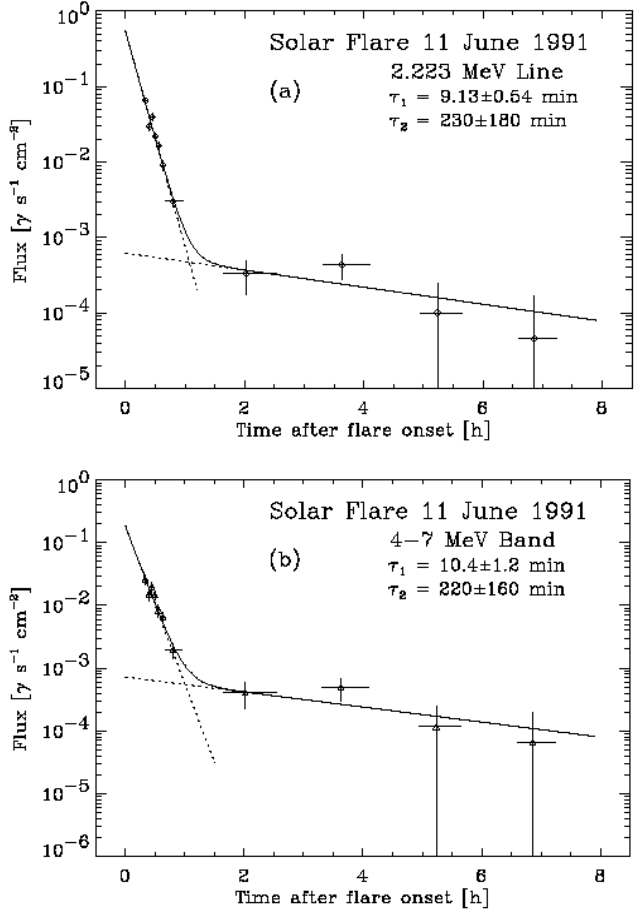}}
\caption{Extended $\gamma$-ray emission as measured by {\sl COMPTEL}
for the GLE event \#51 on
1991-Jun-11 flare in the 2.223 MeV neutron capture line (top)
and the $4-7$ MeV nuclear line flux (bottom). The data have been corrected
for primary and secondary bremsstrahlung. A two-fold exponential decay has
been fitted. The origin of the time axis is 01:56 UT, and the flare onset
reported by GOES. Only data of the extended phase (after 02:13 UT) are
shown (Rank et al.~2001).}
\end{figure}

\subsection{	Extended Particle Acceleration and Trapping 	}

The solar particle release time $t_{SPR}$ as determined by Reames (2009b) 
occurs after the peak time $t_m$ of the hard X-ray production phase in 
92\% (12 out of 13 cases), or after the end time $t_e$ of impulsive hard 
X-rays in 50\% (6 out of 12 non-occulted cases), for the GLE events listed 
in Table 1 and shown in Fig.~1. The question arises whether
we can exclude an interpretation in terms of a flare acceleration site
for such delays. If the flare site shows, besides the impulsive
hard X-ray phase, also prolonged time intervals with hard X-ray and 
gamma-ray emission, either an extended (second-step) acceleration phase
or extended trapping is possible (e.g., Vilmer et al.~1982), 
in which case the particles responsible
for GLE signatures could possibly be accelerated at the coronal flare
site rather than in heliospheric CME shocks. Extended acceleration
can be diagnosed from impulsive bursts in hard X-ray, gamma-ray,
and radio emission after the impulsive phase, while indicators of
extended trapping are: (1) exponential decay of X-ray and gamma-ray
light curves, (2) spectral hardening in hard X-rays (Kiplinger 1995;
Grayson et al.~2009), or (3) type IV continuum emission in microwaves 
(produced by gyrosynchrotron emission 
from trapped high-relativistic electrons). Particle trapping times are
limited by the collisional deflection time, which is for electrons
$$
        t_{trap}(\varepsilon) \lapprox
        t_{defl}(\varepsilon) = 0.95
        \left( {{\varepsilon} \over 100\ {\rm keV}} \right)^{3/2}
        \left( {10^{11}\ {\rm cm}^{-3} \over n_e} \right) \
        \left( {20 \over \ln \Lambda} \right) \quad (s) \ ,
	\eqno(4) 
$$
and a factor of $\approx 60$ longer for ions. Postflare loops usually
expand slowly with height, so that the density drops continuously and
collisional deflection times (and thus trapping times) become progressively
longer. Particle densities in the upper corona drop below 
$n_e \lapprox 10^8$ cm$^{-3}$, which enables trapping times over
several hours. 

The second diagnostic of extended acceleration, the spectral hardening
after the impulsive phase, has recently been demonstrated with RHESSI 
to be a reliable predictor of SEP events (Grayson et al. 2009). From a
dataset of 37 magnetically well-connected flares, 12 out of 18 flares
with spectral soft-hard-harder (SHH) evolution produced SEP events, and none
of the 19 flares without SHH behavior produced SEPs. Three of the 37
analyzed SEP events are GLE events and show all SHH behavior.
This demonstrates a statistically significant correlation between SHH and 
SEP (and GLE) observations.
Since the spectral hardening happens in the hard X-rays sources at the 
solar flare site, as imaged by RHESSI, this link between SHH and SEP is
unexplained in the standard scenario of SEP acceleration in CME-related
shocks during interplanetary propagation.  

Extended gamma-ray emission with exponential decay times
has indeed been observed up to 5-8 hours after the impulsive phase of
the GLE event \#51 on 1991 Jun 11 (Fig.~7) and GLE event \#52 on 
1991 Jun 15 (Kanbach et al.~1993; Rank et al.~1996, 2001). 
Evidence for prolonged acceleration of high-energy protons ($E>100$ MeV) 
accelerated in flares and postflare loop systems, with associated
spectral hardening late in the flare, was also discussed 
for GLE event \#52, 1991 Jun 15, 08:17 UT (Chertok 1995),
for GLE event \#55, 1997 Nov 6, 11:55 UT (Klein and Trottet 2001; 
Murphy et al.~2001; Masuda 2002; Masuda and Sato 2003),
for GLE event \#59, 2000 Jul 14, 10:24 UT (Livshits and Belov 2004; 
Li et al.~2007; Wang 2009),
and for GLE event \#70, 2006 Dec 13, 02:27 UT (Li et al.~2009).
Evidence for extended particle acceleration in the lower corona was also 
demonstrated from microwave gyrosynchrotron emission, requiring high 
magnetic fields near sunspots, e.g., for GLE event \#69, 2005 Jan 20, 
06:48 UT (Grechnev et al.~2008; Masson et al.~2009).
The GLE event \#65 on 2003 Oct 28, 11:05 UT, shows multiple phases
of impulsive particle injections, ending with a gradual phase ($>1$ hr)
of type III bursts that starts 25 min after the first type III bursts  
(Klassen et al.~2005; Miroshnichenko et al.~2005).
While there is ample evidence for extended acceleration of electrons
in flare sites and flare-associated radio burst sources, the situation
is less clear for ions and protons; extended acceleration of ions and 
protons can occur concomitantly in the same flare site, or alternatively 
in propagating CME shocks (e.g., Cliver 1996).

Let us also review the four GLE events listed in Table 1 that
exhibit the largest delays in solar particle release times $t_{SPR}$. 
The GLE event \#56 on 1998 May 02, 13:38 UT, has a delay of 
$t_{SPR}-t_{HXR} = +15.4$ min, at a time when the impulsive flare phase 
is over, but the radio dynamic spectra from Artemis IV show 
type IV continuum emission at frequencies of $\nu \approx 300-700$ MHz 
during 13:42-13:50 UT (see Fig.~8 in Pohjolainen et al.~2001), 
a clear sign of gyro-synchroton radiation from trapped high-relativistic
electrons during the GLE particle release time. 

The GLE event \#58 on 1998 Aug 24, 22:05 UT, has the longest delay
of $t_{SPR}-t_{HXR} = +34.5$ min, but the USAF/ RSTN data 
(http://cdaw.gsfc.nasa.gov/meetings/lws-cdaw2009/data/White/)
show a late increase of 15.4 GHz radio emission after 22:30 UT, which
is an indicator of trapped high-relativistic electrons.

The GLE event \#61 on 2001 Apr 18, 02:12 UT, has a delay
of $t_{SPR}-t_{HXR} = +19.2$ min, but this flare is occulted
by $30^\circ$ behind the limb, which may explain the offset timing
due to the missed hard X-ray and soft X-ray peak. 
In an occulted flare we do not see the prompt hard X-rays from
the footpoints, but only the prompt soft X-rays in coronal heights visible
above the limb. The coronal high-temperature plasma above the limb cools
faster (by conductive cooling) than the bulk of occulted soft X-ray plasma
that gradually fills the postflare loops by chromospheric evporation. Thus, 
the time profile of soft X-ray emission in occulted flares decays faster
than in unocculted flares, and thus we underestimate the full duration of
impulsive hard X-ray emission using the time derivative of soft X-rays.

The GLE event \#66 on 2003 Oct 29, 20:44 UT, has a delay 
of $t_{SPR}-t_{HXR} = +19.5$ min, but the USAF/RSTN data show 
very bursty radio spikes late in the flare during 20:50-20:56 UT 
at 610 MHz, a possible indication of a secondary acceleration phase
producing electron beams in the lower corona.
In summary, we observe in all cases with late GLE solar particle
release times signs of extended particle acceleration or trapping,
which does not exclude that the particles responsible for the GLE 
signatures have been accelerated and temporarily trapped (for
$\approx 10-30$ min) at the solar flare site.

\section{		Conclusions 				 }

We explored here the question whether the largest SEP and GLE events 
that accelerate ions with energies of $\gapprox 1$ GeV could be 
accelerated in solar flare regions, in contrast to the generally
accepted paradigm of acceleration in heliospheric CME shocks. 
We reviewed the {\sl pro and con} aspects from the solar flare site 
that are relevant to answer this question, while the complementary
aspects from CME-associated shocks are discussed in the companion
article by Gang Li. 
The major arguments for a flare-driven acceleration
of GeV particles are: (1a) the coincidence of the particle
release time with the impulsive flare phase (in 50\%),
(2a) the similarity of the impulsive time profiles of the
first GLE component with the impulsive flare phase in
hard X-rays, (3a) observational evidence of maximum
energies $\lapprox 100$ MeV in electron bremsstrahlung,
and (4a) the relatively high magnetic field
and free magnetic energy that is available in flare sites,
compared to the locations of interplanetary CME shocks.
In contrast, major arguments against this interpretation are:
(1b) significant delays of the particle release time with
respect to the impulsive flare phase (in 50\%), and
(2b) the lack of direct observational evidence and 
diagnostics for $\gapprox 1$ GeV protonts and ions in flare sites.
The conclusions are based on observations of
70 GLE events over the last six decades, in particular on the
13 GLE events during the last solar cycle 23 (1998-2006) that provided
excellent new imaging data in gamma rays and hard X-rays (RHESSI),
in soft X-rays and EUV (TRACE, SOHO/EIT), and particle data from IMP,
WIND, and ACE. Our conclusions are:

\begin{enumerate}
\item{Solar particle release times $t_{SPR}$ of GLE-producing particles
	overlap with the impulsive phase of gamma-ray and hard X-ray
	emission in solar flares in 6 out of 12 (non-occulted)cases, and 
	thus the acceleration time of GLE particles is consistent with
	the flare site in 50\% of the cases, taking the full duration of 
	impulsive flare hard X-ray emission ($t_X \approx 3-13$ min)
	into account.}
\item{In the remaining cases, 6 our of 12 occur delayed to the flare peak
	by $\approx 10-30$ min, but observational signatures of extended
	acceleration and/or particle trapping are evident in all strongly 
	delayed cases, and thus all GLE events could potentially be accelerated
	in flare sites. The alternative explanation of delayed second-step
	acceleration in CME-associated shocks cannot be ruled out, however, 
	possibly constituting a secondary gradual GLE component.}
\item{The height of the acceleration region of \lapprox 1 GeV electrons
	and ions depends on the interpretation, being $h \lapprox 0.05$
	solar radii for flare site acceleration (according to electron
	time-of-flight measurements), or $h \approx 2-5$ solar radii
	for CME shock acceleration (according to type II source heights
	and GLE solar particle release delays).}
\item{The magnetic topology at the particle acceleration site is not
	well-known from magnetic modeling or tracing of coronal structures,
	but there is statistical evidence for open-field regions in most
	active regions that provide escape routes for GLE particles.}
\item{The recently discovered strong correlation between the spectral 
	soft-hard-harder (SHH) evolution of solar hard X-rays and SEP events
	poses a new challenge. It is presently unclear how the SHH evolution
	can be explained in the context of the standard scenario in terms
	of SEP acceleration in CME-associated shocks.}
\item{The maximum particle energies observed in solar flares reach up to
	several 100 MeV for electrons and above 1 GeV for ions. The
	required particle energies are $\lapprox 1$ GeV for the
	observed bremsstrahlung continuum at photon energies of
	$\lapprox 10$ MeV, and $0.2-5$ GeV for pion decay radiation
	observed at photon energies of 10 MeV $-$3 GeV, which is
	sufficient to explain GLE detections.}
\item{Energies up to $\lapprox 1$ GeV can be achieved for DC electric
	field acceleration in super-Dreicer fields, for stochastic
	acceleration, and for shock acceleration. Only DC electric field
	acceleration in sub-Dreicer fields can be ruled out, based on
	the large distances of DC fields required and the inconsistency
	with measured electron time-of-flight delays.}
\end{enumerate}
	
What do we need to answer the question of the origin of GeV
particles with more certainty? The velocity dispersion measurements with
IMP, WIND, and ACE helped enormously to narrow down the start window of
solar particle release times, which could be complemented with 
directivity measurements of the particle detector IMPACT on STEREO. 
The most sensitive
hard X-ray and gamma-ray detectors existed on CGRO, but in near future
we have only RHESSI available. However, RHESSI provides excellent imaging
capabilities, which revealed intriguing information on displaced electron
and ion acceleration sites in the solar corona (Hurford et al.~2003, 2006).
In addition we expect also improved magnetic modeling of solar flare regions
that can reveal open-field and closed-field geometries with more certainty,
using nonlinear force-free field (NLFFF) codes (e.g., Bobra et al.~2008;
DeRosa et al.~2009) with 3-D vector magnetograph data from the {\sl Solar
Dynamics Observatory (SDO)}, possibly in conjuction with stereoscopic 3-D
reconstruction of flaring active regions using {\sl STEREO/EUVI}. 

In conclusion, acceleration of GeV particles in flare sites is a possibility
that cannot be firmly ruled out with the current localization capabilities
of energetic particles. Certainly we have evidence for both acceleration in
coronal flare sites and in heliospheric CME shocks, often appearing 
concomitantly, but with different (impulsive vs. gradual) time scales,
relative timing, and charge state characteristics. While one-sided emphasis 
has been given to both, either flares (the ``big flare syndrome''), or CMEs 
(the ``flare myth''; Gosling 1993), there is a consensus now that both flare 
and CME phenomena are part of a common magnetic instability, and that both are
being able to accelerate particles to high energies. The remaining questions
are then mostly what the relative proportions of both components are and
how we can discriminate between them. A preliminary answer is that the
observations are mostly consistent with a flare-associated ``prompt GLE 
component'' and a CME-associated ``delayed GLE component''.

\begin{acknowledgements}
We thank the anonymous referee for pointing out a crucial timing correction,
and Ludwig Klein for complementary comments.
This work resulted from two Coordinated Data Analysis Workshops (CDAW) on
Ground Level Enhancement Events (GLEs), held in January 2009 in Palo Alto,
California, and in November 2009, Huntsville, Alabama, organized by
Nariaki Nitta and Nat Gopalswamy.
\end{acknowledgements}

\section*{References} %REFERENCES
 
\def\ref#1{\par\noindent\hangindent1cm {#1}}

\ref Akimov, V.V. and 33 co-authors 1991, 
	in Proc. 22nd Internat. Cosmic Ray Conference,
	Internat. Union of Pure and Applied Physics (IUPAP), 
	The Institute for Advanced Studies: Dublin, Vol. 3, p. 73.
\ref Akimov, V.V., Belov, A.V., Chertok, I.M., Kurt, V.G., Leikov, N.G.,
	Magun, A., and Melnikov, V.F. 1994a, Proc. Kofu Symposium,
	Kofu, Japan, p.371.
\ref Akimov, V.V., Leikov, N.G., Belov, A.V., Chertok, I.M., Kurt, V.G.,
	Magun, A., and Melnikov, V.F. 1994b, 
	in {\sl High-energy solar phenomena - A new era of spacecraft 
	measurements} (eds. Ryan, J. and Vestrand, W.T.),
 	American Institute of Physics: New York, p.106.
\ref Akimov, V.V., Leikov, N.G., Kurt, V.G., and Chertok, I.M. 1994c, 
	in {\sl High-energy solar phenomena - A new era of spacecraft 
	measurements} (eds. Ryan, J. and Vestrand, W.T.),
 	American Institute of Physics: New York, p.130.
\ref Akimov, V.V., Ambroz, P., Belov, A.V., Berlicki, A., Chertok, I.M.,
	Karlicky, M., Kurt, V.G., Leikov, N.G., Litvinenko, Y.E.,
	Magun, A., Minko-Wasiluk, A., Rompolt, B., and Somov, B.V. 1996, 
	Solar Phys. 166, 107.
\ref Aschwanden, M.J., Kosugi, T., Hudson, H.S., Wills, M.J., and
	Schwartz, R.A. 1996, ApJ 470, 1198. 
\ref Aschwanden, M.J. and Alexander, D. 2001, SP 204, 91.
\ref Aschwanden, M.J. 2002, Space Science Reviews 101, 1.
\ref Aschwanden, M.J. 2004 (1st Edition; 2005 paperback),
        {\sl Physics of the Solar Corona - An Introduction}, Praxis
        Publishing Ltd., Chichester UK, and Springer, New York.
\ref Aschwanden, M.J. and Aschwanden, P.D. 2008, ApJ 674, 530.
\ref Aulanier, G., DeLuca, E.E., Antiochos, S.K., McMullen, R.A. and Golub, L.
 	2000, \apj 540, 1126.
\ref Bieber, J.W., Clem,J., Evenson, P., Pyle, R., Ruffolo, D., 
	Saiz, A., and Wechakama, M. 2008, in Proc. 30th Internat.
	Cosmic Ray Conf. (eds. Caballero et al.), 
	Universidata Nacional Autonoma de Mexico, Mecico, 
	p.229-232.
\ref Bobra, M.G., VanGallegooijen, A.A., and DeLuca, E.E. 2008, \apj 672, 1209.
\ref Bombardieri, D.J., Duldig, M.L., Humble, J.E., and Michael, K.J.
	2008, ApJ 682, 1315. 
\ref Chertok, I.M. 1995, in 24th Internat. Cosmic Ray Conference,
	(eds. Iucci, N. and Lamanna,E.), Internat. Union of Pure and
	Applied Physics, Vol. 4, p.78.
\ref Chertok, I.M., Formichev, V.V., Gnezdilov, A.A., Gorgutsa,R.V.,
	Grechnev, V.V., Markeev, A.K., Nightingale, R.W., and Sobolev, D.E.
	2001, \SP 204, 141.
\ref Chupp, E.L., and Ryan, J.M. 2009, Research in Astron. Astrophys.
	9/1, 11. 
\ref Cliver, E.W. 1996, in {\sl High energy solar physics}, AIP Conf. Proc.
	374, 45. 
\ref Cliver, E.W., Kahler, S.W., Shea, M.A., and Smart, D.F. 1982,
	ApJ 260, 362. 
\ref Cliver, E.W., Kahler, S.W., Cane,H.V., Koomen, M.J., Michels, D.J.,
	Howard, R.A., and Sheeley, N.R.Jr. 1983, \SP 89, 181.
\ref Cliver, E.W. 2006, ApJ 639, 1206.
\ref Dennis, B.R. and Zarro, D.M. 1993, SP 146, 177.
\ref DeRosa, M.L., Schrijver, C.J., Barnes, G., Leka, K.D., Lites, B.W., 
	Aschwanden, M.J., Amari, T., Canou, A., McTiernan, J.M., Regnier, S., 
	Thalmann, J., Valori, G., Wheatland, M.S., Wiegelmann, T., Cheung, M.C.M., 
	Conlon, P.A., Fuhrmann, M., Inhester, B., and Tadesse, T. 2009,
  	\apj 696, 1780. 
\ref Dingus, B.L., Sreekumar, P., Bertsch, D.L., Schneid, E.J.,
	Brazier, K.T.S., Kanbach, G., von Montigny, C., Mayer-Hasselwander, 
	H.A., Lin, Y.C., Michelson, P.F., Nolan, P.L., Kniffen, D.A.,
	Mattox, J.R. 1994, 
	in {\sl High-energy solar phenomena - A new era of spacecraft 
	measurements} (eds. Ryan, J. and Vestrand, W.T.),
 	American Institute of Physics: New York, p.177.
\ref Forbes, T.G. and Priest, E.R. 1995, ApJ 446, 377. 
\ref Forrest, D.J., Vestrand, W.T., Chupp, E.L., Rieger, E., Cooper, J.F.,
	and Share, G.H. 1985, in 19th Intern. Cosmic Ray Conf.,
	NASA Goddard Space Flight Center: Greenbelt, Maryland,
	Vol. 4., p.146.
\ref Gopalswamy, N., Xie, Hl, Yashiro, S., and Usoskin, I. 2010,
	Indian J. Radio and Space Physics 39, 240-248.
\ref Gosling, J.T. 1993,
	JGR 98, A11, p. 18937.
\ref Grayson, J.A., Krucker, S., and Lin, R.P. 2009, \apj 707, 1588. 
\ref Grechnev, V,V, Kurt, V.G., Chertok, I.M., Uralov, A.M., Nakajima, H.,
	Altyntsev, A.T., Belov, A.V., Yushkov, B.Yu., Kuznetsov, S.N.,
	Kashapova, L.K., Meshalkina, N.S., and Prestage, N.P. 2008,
	Solar Phys. 252, 149.
\ref Holman, G.D. 1996, in {\sl High Energy Solar Physics},
        (eds. Ramaty, R., Mandzhavidze, N., and Hua, X.-M.),
	American Institute of Physics: Woodbury, New York,  
	Conf. Proc. 374, p.479.
\ref Hudson, H.S. 1991, BAAS 23, 1064.
\ref Hurford, G.J., Schwartz, R.A., Krucker, S., Lin, R.P., Smith, D.M., 
	and Vilmer, N. 2003, \apj 595, L77.
\ref Hurford, G.J., Krucker, S., Lin, R.P., Schwartz, R.A., Share, G.H., 
	and Smith, D.M. 2006, \apj 644, L93.
\ref Kanbach, G., Bertsch, D.L., Fichtel, C.E., Hartman, R.C., Hunter, S.D.,
	Kniffen, D.A., Kwok, P.W., Lin, Y.C., Mattox, J.R., and
	Mayer-Hasselwander, H.A. 1992, 
        in {\sl EGRET Mission and Data Analysis}, 
   	Technical Report N94-19462 04-89, Max-Planck-Inst. f\"ur Physik 
	und Astrophysik: Munich, p.5.
\ref Kanbach, G, Bertsch, D.L., Fichtel, C.E., Hartman, R.C., Hunter, S.D.,
	Kniffen, D.A., Kwok, P.W., Lin, Y.C., Mattox, J.R. and
	Mayer-Hasselwander, H.A. 1993, AASS 97, 349. 
\ref Karpen, J.T., Antiochos, S.K., and DeVore, C.R. 1995, ApJ 450, 422.
\ref Karpen, J.T., Antiochos, S.K., DeVore, C.R.,a nd Golub, L. 1998, 
	ApJ 495, 491.
\ref Kiplinger, A.L. 1995, \apj 453, 973. 
\ref Klassen, A., Krucker, S., Kunow, H., M\"uller-Mellin, R.,
	Wimmer-Schweingruber, R. 2005, JGR 110, A09S04. 
\ref Klein, K. and Trottet, G. 2001, American Geophysical Union,
	Meeting abstract \#SH31C-09.
\ref Kliem, B., Karlicky, M., and Benz, A.O. 2000, A\&A 360, 715.
\ref Kocharov, L., Klein, K.-L., Saloniemi, O., Kovaltsov, G., and Torsti, J.
	2006, 36th COSPAR Scientific Assembly, Meeting abstract \#2897.
\ref Kundu, M.R. 1965, {\sl Solar Radio Astronomy}, Wiley: New York.
\ref Kurt, V.G., Akimov, V.V., and Leikov, N.G. 1996, 
        in {\sl High Energy Solar Physics},
        (eds. Ramaty, R., Mandzhavidze, N., and Hua, X.-M.),
	American Institute of Physics: Woodbury, New York,  
	Conf. Proc. 374, p.237.
\ref Kuznetsov, S.N., Kurt, V.G., Yushkov, B.Y., Myagkova, I.N.,
	Kudela, K., Kassovicova, J., and Slivka, M. 2006,
	Contrib. Astron. Obs. Skalnat\'e Pleso 36, 85.  
\ref Li, J., Tang, Y.H., Dai, Y., Zong, W.G., and Fang, C. 2007,
	A\&A 461, 1115. 
\ref Li, J., Dai, Y., Vial, J.C., Owen, C.J., Matthews, S.A., Tang, Y.H.,
	Fang, C., Fazakerley, A.N. 2009, A\&A 503, 1013.
\ref Litvinenko, Y.E. 1996, \apj 462, 997.
\ref Litvinenko, Y.E. 2003, Solar Phys. 216, 189.
\ref Livshits, M.A. and Belov, A.V. 2004, Astronomy Reports 48/8, 665.
\ref Lopate, C. 2006, {\sl ``Solar Eruptions and Energetic Particles''},
	(eds., Gopalswamy, N., Mewaldt,R., and Torsti, J.),
	Geophysical Monograph Series Vol. 165, p.283. 
\ref Mandzhavidze, N. and Ramaty, R. 1992, \apj 389, 739.
\ref Martirosyan, H. and Chilingarian, A. 2005, Proc. 29th Internat.
	Cosmic Ray Conf. (eds. Acharya, B.S. et al.), Tata Institute
	of Fundamental Research: Mumbai, Vol.2, p.285. 
\ref Masson, S., Klein, K.-L., B\"utikover, R., Fl\"uckiger, E.O.,
	Kurth, V., Yushkov,  B., and Krucker, S. 2009,
	Solar Phys. 257, 305.
\ref Masuda, S. 2002, in {\sl Multi-wavelength observations of coronal
	structure and dynamics - Yohkoh 10th Anniversary Meeting},
	(eds. Martens, P.C.H. and Cauffman, D.), COSPAR Coll. Series,
	Elsevier, p.351.
\ref Masuda, S., and Sato, J. 2003, Adv. Space Res. 32/12, 2455.
\ref Matthi\"a, D., Heber, B., Reitz, G., Meier, M., Sihver, L., Berger, T.,
	and Herbst K. 2009, J. Geophys. Res. 114/A8, CiteID A08104. 
\ref McCracken, K.G. and Moraal, H. 2008, 
	in Proc. 30th Internat.  Cosmic Ray Conf. (eds. Caballero et al.), 
	Universidata Nacional Autonoma de Mexico, Mecico, p.269. 
\ref McCracken, K.G., Moraal, H., and Stoker, P.H. 2008, JGR 113, A12101.
\ref Miller, J.A., Cargill, P.J., Emslie, A.G., Holman, G.D., Dennis, B.R., 
	LaRosa, T.N., Winglee, R.M., Benka, S.G., and Tsuneta, S. 1997, 
	JGR 102/A7, 14631.
\ref Miroshnichenko, L.I., Klein, K.-L., Trottet, G., Lantos, P.,
	Vashenyk, E.V., Balabin, Y.V., and Gvozdevsky, B.B. 2005, 
	JGR 110/A11, CiteID A11s90. 	
\ref Moraal, H., McCracken, K.G., and Stoker, P.H. 2008,
	in Proc. 30th Internat.  Cosmic Ray Conf. (eds. Caballero et al.), 
	Universidata Nacional Autonoma de Mexico, Mecico, p.265. 
\ref Moraal, H., and McCracken, K.G. 2008, Space Science Rev. (this issue),
	(in press).	
\ref Murphy, R.J., Share, G.H., Schwartz, R.A., Yoshimori, M., Suga, K.,
	Nakayama, S., Takeda, H. 2001, American Geophysical Union,
	Meeting abstract \#SP42A-10. 
\ref Neupert, W.M. 1968, ApJ 153, L59.
\ref Pohjolainen, S., Maia, D., Pick, M., Vilmer, N., Khan, J.I., Otruba, W.,
	Warmuth, A., Benz, A., Alissandrakis, C., and Thompson, B.J.
	2001, \apj 556, 421. 
\ref Ramaty, R. and Mandzhavidze, N. 1994, 
	in {\sl High-energy solar phenomena - A new era of spacecraft 
	measurements} (eds. Ryan, J. and Vestrand, W.T.),
 	American Institute of Physics: New York, p.26.
\ref Rank, G., Bennett, K., Bloemen, H., Debrunner, H., Lockwood, J.,
	McConnell, M., Ryan, J., Sch\"onfelder, V., and Suleiman, R.
	1996, in {\sl High Energy Solar Physics}, AIP Conf. Proc. 374,
	(eds. Ramaty, R., Mandzhavidze, N., and Hua, X.-M.),
	American Institute of Physics: New York, p.219. 
\ref Rank, G., Ryan, J., Debrunner, H., McConnell, M., and Schoenfelder, V.
 	2001, A\&A 378, 1046.
\ref Reames, D.V., Richardson, I.G., and Wenzel, K.P. 1992, \apj 387, 715.
\ref Reames, D.V. 2002, ApJ 571, L63.
\ref Reames, D.V. 2009a, ApJ 693, 812.
\ref Reames, D.V. 2009b, ApJ 706, 844.
\ref Schrijver, C.J. and DeRosa, M.L. 2003, \SP 212, 165.  
\ref Shibata, K. and Tanuma, 2001, Earth, Planets and Space 53, 473. 
\ref Shumilov, O.I., Kasatkina, E.A., Turyansky, V.A., Kyro, E., Kivi, R.
	2003, Adv. Space Res. 31/9, 2157. 
\ref Simnett, G.M. 2006, A\&A 445, 715.  
\ref Simnett, G.M. 2007, (Erratum) A\&A 472, 309.  
\ref Simnett, G.M. and Roelof, E.C. 2005a, 29th Internat Cosmic Ray Conf.
	1, 233. 
\ref Simnett, G.M. and Roelof, E.C. 2005b, AGU Meeting abstract \#SH23A-0319. 
\ref Sturrock, P.A. 1966, Nature 5050, 695.
\ref Torsti, J., Riihonen, E., and Kocharov, E. 2004, ApJ 600, L83. 
\ref Trottet, G. 1994, 
	in {\sl High-energy solar phenomena - A new era of spacecraft 
	measurements} (eds. Ryan, J. and Vestrand, W.T.),
 	American Institute of Physics: New York, p.3.
\ref Vashenyuk, E.V., Balabin, Y.V., and Gvozdevsky, B.G. 2003,
	28th Internat. Cosmic Ray Conf., p.3401.
\ref Vashenyuk, E.V., Balabin, Y.V., and Gvozdevsky, B.G. 2011,
	Astrophys. Space Sci. Transactions, in Proc.
	22th Internat. Cosmic Ray Conf., Turku, Finland, (subm).
\ref Vilmer, N., Kane, S.R., and Trottet, G. 1982,
 	A\&A 108, 306.	
\ref Wang, H., Gary, D.E., Zirin, H., Schwartz, R.A., Sakao, T., Kosugi, T.,
	and Shibata, K., 1995, ApJ 453, 505.
\ref Wang, R.G. and Wang, J.X. 2006,
	36th COSPAR Scientific Assembly, Meeting abstract \#1856.
\ref Wang, R.G. 2009, Astroparticle Physics 31/2. 149. 
\ref Watanabe, K., Murphy, R.J., Share, G.H. et al. 2008, 
	in Proc. 30th Internat.  Cosmic Ray Conf. (eds. Caballero et al.), 
	Universidata Nacional Autonoma de Mexico, Mecico, p.41. 
\ref Yan, Y. and Huang, G. 2003, Space Science Rev. 107, 111.
\ref Zank, G.P., Rice, W.K.M., and Wu, C.C. 2000, JGR 105/A11, 25079. 

\end{document}